\documentstyle[aps,prl,epsfig]{revtex}

\draft

\begin{document}

\title{\bf Trapping and cooling single 
atoms with far-off resonance intracavity
doughnut modes}

\author{Stefano Pirandola, David Vitali, and Paolo Tombesi}

\address{
INFM, Dipartimento di Fisica,
Universit\`a di Camerino,
I-62032 Camerino, Italy}

\date{Received: \today}

\maketitle 
\widetext

\begin{abstract}
We investigate cooling and trapping of single atoms inside an optical cavity
using a quasi-resonant field and a far-off resonant mode of the
Laguerre-Gauss type. The far-off resonant doughnut mode provides an
efficient trapping in the case when it shifts the atomic internal ground and
excited state in the same way, which is particularly useful for quantum
information applications of cavity quantum electrodynamics (QED) systems.
Long trapping times can be achieved, as shown by full 3-D simulations of the
quasi-classical motion inside the resonator.
\end{abstract}


\widetext

\section{Introduction}

Cavity quantum electrodynamics (QED) is a powerful tool for the
deterministic control of atom-photon interactions at the quantum level. In
fact, the strong confinement allows to achieve the strong coupling regime
where single quanta can profoundly affect the atom-cavity dynamics \cite
{berm}. Many experiments have now reached this regime and interesting
phenomena have been demonstrated, as a quantum phase gate \cite{turch}, the
Fock state generation in a cavity field \cite{varcoe}, and quantum
nondemolition detection of a single cavity photon \cite{nog}. Moreover,
cavity QED offers unique advantages for quantum communication and quantum
information processing applications. In fact, atoms may act as quantum
memories, while photons are flexible transporters of quantum information,
and quantum networks of multiple atom-cavity systems linked by optical
interconnects have been already discussed in the literature \cite{pell}. The
primary technical challenge on the road toward these applications is to trap
individual neutral atoms within a high-finesse cavity for a reasonably long
time. Recent experiments \cite{hood,pinkse} have already succeeded in
trapping single atoms inside an optical cavity driven at the few photon
level, just using the strong coupling with the cavity QED mode for both
cooling and trapping. However, the scheme of Refs.~\cite{hood,pinkse} is not
entirely suitable for quantum communication purposes because it has limited
operation flexibility and provides short trapping times (order of hundreds
of microseconds). In fact, it employs a single cavity mode, while it is
preferable to have an additional trapping mechanism which does not interfere
with the cavity QED interactions able to provide the atom-photon
entanglement needed for the manipulation of quantum information. With these
respect, another experiment has already demonstrated significant
trapping times ($\sim 28$ ms) of single Cs atoms within a cavity, employing
an additional far-off resonance trapping (FORT) mode \cite{Ye-exp}. Using
the fact that, in the strong coupling regime, the trajectory of an
individual atom can be monitored in real time by the quasi-resonant cavity
QED field \cite{ACM}, the FORT beam can be turned on as soon as the atom
enters the cavity in order to increase its trapping time.

Several mechanisms for cooling inside an optical resonator have been already
discussed in the literature \cite{older,ritsch,vule,doherty}, involving
either cavity mode driving, or direct atom driving via a classical laser
field from the side, or even active feedback on atomic motion as in 
\cite{feedback}. However, only the recent paper
by van Enk {\it et al.} \cite{vanenk} has discussed in detail the effects of
an additional FORT beam on the cooling and trapping dynamics and its
interplay with the quasi-resonant cavity QED field. Here we shall consider a
situation analogous to that of Ref.~\cite{vanenk}, even though we shall
extend our study to new FORT mode configurations. We have chosen a parameter
region corresponding to the weak driving limit, with an empty-cavity mean
photon number $N_{e}\simeq 0.01$. We have performed full 3-D numerical
simulations of the quasi-classical atomic motion, including the effects of
spontaneous emission and dipole-force fluctuations.

In two-level systems, red-detuned FORT beams shift the atomic excited state $%
|e\rangle $ {\em up} and the ground state $|g\rangle $ {\em down} by the
same quantity, and this is the most common situation, studied in great
detail in \cite{vanenk}. However, the most interesting situation for quantum
information processing applications is when both levels are shifted {\em down%
} by the FORT beam: in this case, excited and ground state atoms are trapped
in the same position, and this greatly simplifies the quantum manipulation
of the internal state. In fact, the most flexible situation for quantum
information processing is having a trapping mechanism independent of the
atomic internal state. This configuration can be realized by using a FORT
that is red-detuned in such a way that the excited state is relatively
closer to resonance with a higher-lying excited state than with the ground
state. We shall discuss in detail both situations (equal or opposite optical
Stark shifts), and we shall find the interesting result that in the case of
equal Stark shifts for ground and excited state, the use of doughnut modes,
that is, higher-order Gauss-Laguerre modes, as FORT mode, is able to
increase significantly the trapping time within the cavity.

The paper is organized as follows. In Section II we describe the physics
of an atom trapped in an optical potential and strongly
interacting with a cavity QED field. We also discuss the changes introduced
by the FORT doughnut mode. In Section III, applying the general approach
developed in Ref.~\cite{cohen} to a cavity mode configuration, we discuss
the conditions under which the center-of-mass motion of the atom can be
adiabatically separated from the internal and cavity mode dynamics, and
treated in a quasi-classical way. The corresponding 3-D Fokker-Planck
equation for the phase-space atomic motion will be derived. In Section IV,
the results of the numerical simulations of the corresponding stochastic
differential equations will be presented in detail, and in Section V these
results will be discussed for both equal and opposite energy shifts of the $%
|e\rangle $ and $|g \rangle $ states. Section VI is for concluding remarks.

\section{The physical problem}

\begin{figure}[h]
\centerline{\epsfig{figure=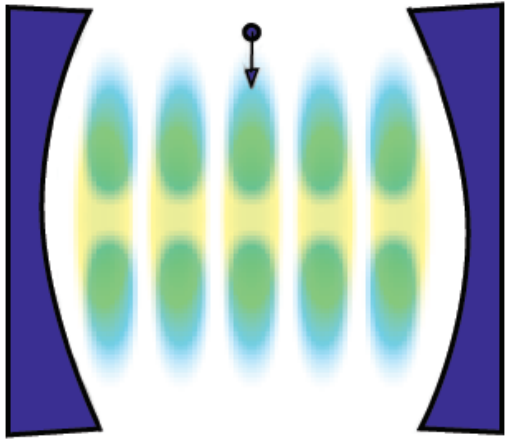,width=3.0in}}
\caption{ Schematic description of the system. Atoms falling from a
magneto-optical trap are trapped within a cavity with a fundamental
Gaussian quantized
resonant mode (yellow) and an intense classical doughnut FORT mode
(light blue).}
\label{appa}
\end{figure}

We consider a two level atom coupled to a quantized cavity mode and to an
additional classical red-detuned FORT beam, coinciding with another
longitudinal mode of the cavity, with a wavelength $\lambda_{S}$ longer than
that of the quasi-resonant cavity mode, $\lambda_{g}$. The common situation
is to consider lowest order Gaussian modes for both fields \cite{vanenk},
having their maximum intensity along the cavity axis. Here we shall consider
a different situation, where the FORT mode is a higher-order Gauss-Laguerre
mode, the so-called doughnut mode, having its maximum intensity at a nonzero
radial distance from the cavity axis. This means that the atoms are trapped
out of the cavity axis (see Fig.~\ref{appa} for a schematic description of
the system). At first sight, this choice may look not optimal, because in
this case the coupling with the fundamental
Gaussian cavity QED field responsible for
cooling is smaller. Nonetheless, we shall see that this choice is convenient
in the case when the classical FORT mode shifts {\em down} both excited and
ground state, which is the most interesting case for quantum information
processing applications.

We recall that the Laguerre-Gauss modes LG$_{pm}$ are the solutions of the
paraxial Helmoltz equation in cylindrical coordinates $(\rho ,\vartheta ,x$) 
\cite{siegman}. In this paper we consider the doughnut modes with radial
index $m\neq 0$ and azimuthal index $p=0$, whose intensity is given by 
\begin{equation}
I_{0m}(\rho ,x)=4P\frac{2^{m+1}}{\pi m!}\frac{\rho ^{2m}}{W^{2(m+1)}(x)}\exp
\left\{ -\frac{2\rho ^{2}}{W^{2}(x)}\right\} \sin ^{2}(k_{S}x),  \label{LG0m}
\end{equation}
where $P$ is the power, $k_{S}=2\pi /\lambda _{S}$ the wavenumber, and $W(x)$
the beam radius. The doughnut mode radius is given by the position of the
radial maximum, given by $\rho _{\max }=W(x)\sqrt{m/2}$. We can simplify the
description assuming a nearly planar cavity, so that $W(x)\sim W_{S}$. The
classical, red-detuned FORT mode induces AC Stark shifts $\Delta E_{g}$ on
the ground state $|g\rangle $ and $\Delta E_{e}$ on the excited state 
$|e\rangle $ \cite{FORT}. We can distinguish two different situations: (a) 
$|g\rangle $ is shifted down and $|e\rangle $ is shifted up, which happens in
the common situation where only the $e\leftrightarrow g$ transition is close
to the red detuned FORT mode. In this case the two shifts are opposite. (b)
Both $|g\rangle $ and $|e\rangle $ are shifted down, which happens when $%
|e\rangle $ is closer to resonance with a higher-lying excited level than
with the level $|g\rangle $ \cite{frequencies},\cite{singapore},\cite{Buck}.
We shall study both situations and in case (b) we shall assume that the
various detunings can be chosen so that the two levels are shifted down by
the same quantity. Furthermore, to simplify the comparison between the two
cases, we shall assume that the two situations occur with the same
wavelength $\lambda _{S}$.

Considering a frame rotating at the probe driving frequency $\omega_{p}$,
the Hamiltonian of the system can be written as 
\begin{equation}
H=\frac{|\vec{P}|^{2}}{2M}+ \hbar\omega_{ap}\sigma^{\dagger}\sigma+\hbar
\omega_{gp}a^{\dagger}a+V(\vec{R})+\hbar\left(Ea^{\dagger}+E^{\ast} a\right),
\label{Jaynes-C}
\end{equation}
where $\omega_{ap}=\omega_{eg}-\omega_{p}$ is the detuning of the the atomic
resonance from the probe frequency, $\omega_{gp}=\omega_{g} -\omega_{p}$ is
the detuning of the cavity QED mode with annihilation operator $a$, $\sigma
= |g\rangle \langle e|$, $E$ is the cavity driving rate, and $\vec{R}$, $%
\vec{P}$ are the position and momentum vector operators of the atom, having
a mass $M$. The interaction potential $V(\vec{R})$ describes the interaction
between the internal atomic levels, the cavity modes and the atomic
center-of-mass motion, and it is given by the coupling with the quantized
cavity mode and with the FORT doughnut mode. Making the usual dipole and
rotating wave approximations, this interaction term can be written 
\begin{equation}
V(\vec{R})=\hbar g(\vec{R})\left( a^{\dagger}\sigma+\sigma^{\dagger}a\right)
+V^{S}(\vec{R}),  \label{potential}
\end{equation}
where 
\begin{equation}
g(\rho,x)=g_{0}\sin(k_{g}x)\exp\left\{-\frac{\rho^{2}}{W_{g}^{2}}\right\}
\label{modo-cavita}
\end{equation}
is the space-dependent Rabi frequency due to the coupling with the quantized
mode, and the second term $V^{S}(\vec{R})$ describes the effect of the Stark
shifts induced by the FORT mode, assuming the following form in the two
cases, (a) and (b): 
\begin{eqnarray}
\text{case (a) \ \ \ }V^{S}(\vec{R})&=&\hbar S(\vec{R})(\sigma^{\dagger}
\sigma-\sigma\sigma^{\dagger})  \label{Stark-a} \\
\text{case (b) \ \ \ }V^{S}(\vec{R})&=&-\hbar S(\vec{R}).  \label{Stark-b}
\end{eqnarray}
The frequency ac-Stark shift is generally given by $S(\rho,x)=\alpha
I_{0m}(\rho,x)/4\hbar $ \cite{Savard,gardiner}, where $\alpha$ is the atomic
polarizability and $I_{0m}(\rho,x)$ is the FORT intensity of Eq.~(\ref{LG0m}%
), so that one can write 
\begin{equation}
S(\rho,x)=S_{0}\rho^{2m}\sin^{2}(k_{S}x)\exp\left\{-\frac{2\rho^{2}} {%
W_{S}^{2}}\right\} ,  \label{ciambella}
\end{equation}
with $S_{0}=2^{m+1}P\alpha /\pi \hbar W_{S}^{2m+2}$. We have generally
considered different waists $W_{g}$ and $W_{S}$ for the two modes, but they
practically coincide in typical situations.

The FORT mode provides the main trapping mechanism. This means that the atom
will be trapped around the anti-nodes of the red-detuned FORT field, because 
$S_{0} > g_{0}$ and the cavity QED field is weakly driven. However, due to
the different wavelengths, one does not have a periodic situation, and the
atom feels a different cavity QED coupling in different wells. In the
experiment of Ref.~\cite{Ye-exp} the cavity length $L$ is such that $2L= 104
\lambda_{g}=102 \lambda_{S}$. To simplify our simulation we have however
chosen $L=16 \lambda_{g}=15\lambda_{S}$, as it has been done also in \cite
{vanenk}. This is equivalent to choose a fictitious larger value for $%
\lambda_{S}$ which however does not modify the essential physics of the
problem, because the FORT mode is in any case far-off resonance. With this
choice we consider only $30$ potential wells in the cavity, only $8$ of
which are quantitatively different (see Ref.~\cite{vanenk}).

Dissipation, diffusion and all non-conservative effects appear due to
spontaneous emission and cavity losses. The quantum evolution of the
atom-cavity system is therefore described by a master-equation for the
atom-cavity density operator $\rho$ \cite{cohen} 
\begin{eqnarray}
\frac{d\rho}{dt} & =& -\frac{i}{\hbar}\left[ H,\rho\right] +\kappa\left(
2a\rho a^{\dagger}-a^{\dagger}a\rho-\rho a^{\dagger}a\right) -\gamma\left(
\sigma^{\dagger}\sigma\rho+\rho\sigma^{\dagger}\sigma\right)  \nonumber \\
&+& \frac{3\gamma}{4\pi}\int d^{2}\hat{k}~D(\hat{k}\cdot\hat{x})e^{-i\vec{k}%
\cdot\vec{R}}\sigma\rho \sigma^{\dagger}e^{i\vec{k}\cdot\vec{R}}
\label{master}
\end{eqnarray}
where $\kappa$ is the cavity damping rate and $\gamma$ is spontaneous
emission decay rate. The last term describes the effects of atomic recoil,
with $\hat{k}$ giving the direction of emitted photon and 
$D(\hat{k}\cdot\hat{x})=(1+(\hat{k}\cdot\hat{x})^{2})/2$ 
describing the angular pattern of
dipole radiation \cite{Javanainen}.

\section{Quasi-classical description of the atomic motion}

The internal and cavity dynamics is governed by the detunings, $\omega_{ap}$
and $\omega_{gp}$, the spontaneous emission rate $\gamma$, the driving rate $%
E$, the cavity damping rate $\kappa$, the FORT shift $S_{0}$, and the
coupling constant $g_{0}$. In the parameter region relevant for current
experiments, this dynamics is much faster than the atomic motional dynamics,
especially for heavy atoms as Cs \cite{Ye-exp} or Rb \cite{pinkse}.
Therefore, internal and cavity dynamics can be adiabatically eliminated in
order to obtain a reduced effective evolution equation for the motional
degrees of freedom only. This adiabatic elimination can be made in a
systematic way by adapting the general approach developed for free space
cooling situations in Ref.~\cite{cohen} to the present cavity scheme (see
also \cite{doherty}). One starts from the evolution equation of the Wigner
operator obtained by performing the Wigner transform only on the motional
Hilbert space, and writing this equation as a Taylor expansion in terms of
two small parameters of the problem. Denoting with $k \simeq k_{g}\simeq k_S
\simeq k_a = \omega_{eg}/c$ the typical wavenumber of the problem, we have
that one small parameter is 
\begin{equation}
\epsilon_{1}=\frac{\hbar k }{\Delta p} \ll 1,  \label{eps1}
\end{equation}
showing that the motional state is characterized by a momentum spread $%
\Delta p$ much larger than the momentum kicks felt by the atom during any
photon emission or absorption. The second small parameter is 
\begin{equation}
\epsilon_{2}\simeq \frac{ k \Delta p }{M \gamma } \simeq \frac{ k \Delta p }{%
M \kappa }\ll 1,  \label{eps2}
\end{equation}
which instead shows that the Doppler shift associated with the momentum
spread is small with respect to the atomic and cavity linewidths. The two
conditions set a lower and an upper bound for the momentum spread of the
atom, which, thanks to the Heisenberg inequality, put also a lower and an
upper bound for its position spread. These bounds allow to describe the
atomic motion in classical terms, because the atom is sufficiently localized
in phase space to make it possible to describe its motion in terms of
trajectories, while at the same time the phase space spread always remains
larger than $\hbar$ (see \cite{ACM,ritsch}). The crucial point is that the
two conditions (\ref{eps1}) and {\ref{eps2}) must be consistent and this
happens when 
\begin{equation}
\frac{\hbar^{2}k^{2}}{2M} \ll \hbar \gamma, \hbar \kappa.  \label{condi}
\end{equation}
This means supplementing the well known necessary condition for laser
cooling in free space, $\hbar^{2}k^{2}/2M \ll \hbar \gamma$ (the atom has to
be still in resonance after spontaneously emitting) \cite{cohen}, with an
analogous condition for the exchange of cavity photons. }

For small parameters $\epsilon_{1}$ and $\epsilon_{2}$, the atomic motion is
much slower than the internal one, and the adiabatic elimination of the
latter is obtained if the atomic kinetic energy term is neglected in the
equation for the Wigner operator, which then effectively factorizes into a
product of a stationary state $\eta\left( \vec{r}\right) $ for the cavity
and internal degrees of freedom evaluated at the fixed atomic position $\vec{%
r}=(\rho,\theta,x)$, and a real-valued motional Wigner function $f(\vec{r},%
\vec{p},t)$. The stationary state $\eta\left( \vec{r}\right) $ satisfies the
steady-state Bloch equation 
\begin{eqnarray}
L_{Bloch}(\vec{r})\eta(\vec{r}) & \equiv &-i\left[ \omega_{ap}\sigma
^{\dagger}\sigma+\omega_{gp}a^{\dagger}a+\left( Ea^{\dagger}+E^{\ast
}a\right) +\frac{V(\vec{r})}{\hbar},\eta(\vec{r})\right]  \nonumber \\
&+&\kappa\left( 2a\eta(\vec{r})a^{\dagger}-a^{\dagger}a\eta(\vec{r}) -\eta(%
\vec{r})a^{\dagger}a\right)  \nonumber \\
& +& \gamma\left( 2\sigma\eta(\vec{r})\sigma^{\dagger}-\sigma^{\dagger}
\sigma\eta(\vec{r})-\eta(\vec{r})\sigma^{\dagger}\sigma\right) =0,
\label{LBloch}
\end{eqnarray}
where $L_{Bloch}(\vec{r})$ comes from a zeroth-order expansion of the master
equation (\ref{master}) in the small parameters $\epsilon_{1}$ and $%
\epsilon_{2}$ of Eqs.~(\ref{eps1}) and (\ref{eps2}).

The resulting equation for the Wigner function $f(\vec{r},\vec{p},t)$ is of
Fokker-Planck type and is given by \cite{doherty,cohen} 
\begin{eqnarray}
\frac{\partial}{\partial t}f(\vec{r},\vec{p},t) & =& -\frac{\vec{p}}{M} \cdot%
\frac{\partial}{\partial\vec{r}}f(\vec{r},\vec{p},t)-\frac{\partial }{%
\partial\vec{p}}f(\vec{r},\vec{p},t)\cdot\vec{\phi}(\vec{r})  \nonumber \\
& +&\hbar^{2}k_a^{2}\gamma\left\langle \sigma^{\dagger}\sigma\right\rangle (%
\vec{r}) \sum_{i}E_{ii}\frac{\partial^{2}}{\partial p_{i}^{2}}f(\vec{r},%
\vec {p},t)+ \sum_{ij}\eta_{ij}(\vec{r})\frac{\partial^{2}}{\partial
p_{i}\partial r_{j}}f(\vec{r},\vec{p},t)  \nonumber \\
& +& \sum_{ij}D_{ij}(\vec{r})\frac{\partial^{2}}{\partial p_{i}\partial p_{j}%
}f(\vec{r},\vec{p},t)+ \sum_{ij}\Gamma_{ij}(\vec{r})\frac{\partial}{\partial
p_{i}}(p_{j}f(\vec {r},\vec{p},t)) .  \label{Fokker-P}
\end{eqnarray}
All the coefficients of this Fokker-Planck equation depend upon average
values and correlation functions of the internal and cavity degrees of
freedom, evaluated on the stationary state at fixed atomic position $%
\eta\left( \vec{r}\right) $ of Eq.~(\ref{LBloch}). An example is provided by
the average value $\left\langle \sigma^{\dagger}\sigma\right\rangle (\vec{r}%
) $ appearing in Eq.~(\ref{Fokker-P}) in the diffusion term due to
spontaneous emission, which also depends upon the diagonal matrix $E_{ii}$
given by $E_{xx}=2/5$, $E_{yy}=E_{zz}=3/10$ \cite{doherty,Javanainen}. The
other atom-cavity quantities determining the Fokker-Planck equation
coefficients are $\Phi\equiv a^{\dagger}\sigma+\sigma^{\dagger}a $ and $%
\Psi\equiv\sigma^{\dagger}\sigma-\sigma\sigma^{\dagger}$, whose expectation
value determines the mean dipole force acting on the atom $\vec{\phi}(\vec{r}%
)$. In fact, 
\begin{equation}
\vec{\phi}(\vec{r})= \left\{\matrix{-\hbar\frac{\partial}{\partial\vec{r}}g(%
\vec {r})\left\langle \Phi\right\rangle (\vec{r})
-\hbar\frac{\partial}{\partial \vec{r}}S(\vec{r})\left\langle
\Psi\right\rangle (\vec{r})& {\rm case \; (a)} \cr
-\hbar\frac{\partial}{\partial\vec{r}}g(\vec{r})\left\langle
\Phi\right\rangle (\vec{r}) +\hbar\frac{\partial}{\partial\vec{r}}S(\vec{r})
& {\rm case \; (b)} \cr} \right.,  \label{forza}
\end{equation}
which is a sum of the quantized mode contribution and the FORT mode
contribution. This is no more true for the friction matrix $\Gamma_{ij}(\vec{%
r})$ and the diffusion matrix $D_{ij}(\vec{r})$, which are bilinear
functionals of the dipole force operator and assume different forms in the
two cases $(a)$ and $(b)$. In fact, in case $(b)$ of equal shifts, the FORT
mode simply adds a conservative potential, independent of the internal
atomic state, giving no contribution to friction and diffusion (provided
that the FORT beam intensity fluctuations are negligible, see \cite
{Savard,gardiner}). Therefore we can write 
\begin{equation}
\Gamma_{ij}(\vec{r})=\left\{\matrix{\Gamma_{ij}^{gg}(\vec{r})+
\Gamma_{ij}^{gS}(\vec{r})+\Gamma_{ij}^{SS}(\vec{r})& {\rm case \;(a)} \cr
\Gamma_{ij}^{gg}(\vec{r}) & {\rm case \;(b)} \cr} \right.,  \label{friction}
\end{equation}
\begin{equation}
D_{ij}(\vec{r})=\left\{\matrix{D_{ij}^{gg}(\vec{r})+D_{ij}^{gS}(\vec{r})+
D_{ij}^{SS}(\vec{r})& {\rm case (a)} \cr D_{ij}^{gg}(\vec{r}) & {\rm case
(b)} \cr} \right.,  \label{diffusione}
\end{equation}
where the quantized mode contribution ($gg$), the FORT mode
contribution ($SS$), and the cross contribution ($gS$) have
been singled out. These terms can be then written as 
\begin{eqnarray}
\Gamma_{ij}^{gg}(\vec{r}) & = & \frac{\hbar}{M}\frac{ \partial g}{\partial
r_{i}} (\vec{r}) \frac{ \partial g}{\partial r_{j}} (\vec{r}) \chi^{gg}(\vec{%
r})  \label{fr-gg} \\
\Gamma_{ij}^{gS}(\vec{r}) & = & \frac{\hbar}{M}\frac{ \partial g}{\partial
r_{i}}(\vec{r}) \frac{ \partial S}{\partial r_{j}}(\vec{r})\chi^{gS}(\vec{r}%
)+ \frac{\hbar}{M}\frac{ \partial S}{\partial r_{i}}(\vec{r}) \frac{
\partial g}{\partial r_{j}}(\vec{r})\chi^{Sg}(\vec{r})  \label{fr-gS} \\
\Gamma_{ij}^{SS} (\vec{r}) & = & \frac{\hbar}{M}\frac{ \partial S}{\partial
r_{i}} (\vec{r}) \frac{ \partial S}{\partial r_{j}} (\vec{r}) \chi^{SS} (%
\vec{r}),  \label{fr-SS}
\end{eqnarray}
and 
\begin{eqnarray}
D_{ij}^{gg}(\vec{r}) & = & \hbar^{2}\frac{ \partial g}{\partial r_{i}} (\vec{%
r}) \frac{ \partial g}{\partial r_{j}} (\vec{r}) \xi^{gg}(\vec{r})
\label{diff-gg} \\
D_{ij}^{gS}(\vec{r}) & = & \hbar^{2}\frac{ \partial g}{\partial r_{i}}(\vec{r%
}) \frac{ \partial S}{\partial r_{j}}(\vec{r})\xi^{gS}(\vec{r})+ \hbar^{2}%
\frac{ \partial S}{\partial r_{i}}(\vec{r}) \frac{ \partial g}{\partial r_{j}%
}(\vec{r})\xi^{Sg}(\vec{r})  \label{diff-gS} \\
D_{ij}^{SS}(\vec{r}) & = & \hbar^{2}\frac{ \partial S}{\partial r_{i}} (\vec{%
r}) \frac{ \partial S}{\partial r_{j}} (\vec{r}) \xi^{SS}(\vec{r}) ,
\label{diff-SS}
\end{eqnarray}
where 
\begin{eqnarray}
\chi^{gg}(\vec{r}) & = & i\int_{0}^{\infty}d\tau\tau\left\langle \left[
\Phi\left( \tau\right) ,\Phi\left( 0\right) \right] \right\rangle
\label{CHI-gg} \\
\chi^{gS}(\vec{r}) & =& i\int_{0}^{\infty}d\tau\tau\left\langle \left[
\Phi\left( \tau\right) ,\Psi\left( 0\right) \right]\right\rangle
\label{CHI-gS} \\
\chi^{Sg}(\vec{r}) & =& i\int_{0}^{\infty}d\tau\tau\left\langle \left[
\Psi\left( \tau\right) ,\Phi\left( 0\right) \right] \right\rangle
\label{gS-CHI} \\
\chi^{SS} (\vec{r}) & =& i\int_{0}^{\infty}d\tau\tau\left\langle \left[
\Psi\left( \tau\right) ,\Psi\left( 0\right) \right] \right\rangle,
\label{CHI-SS}
\end{eqnarray}
and 
\begin{eqnarray}
\xi^{gg}(\vec{r}) & =& \int_{0}^{\infty}d\tau\left[ \frac{1}{2}\left\langle
\left\{ \Phi\left( \tau\right) ,\Phi\left( 0\right) \right\} \right\rangle
-\left\langle \Phi\right\rangle ^{2}\right]  \label{CSI-gg} \\
\xi ^{gS}(\vec{r}) &=&\int_{0}^{\infty }d\tau \left[ \frac{1}{2}\left\langle
\left\{ \Phi \left( \tau \right) ,\Psi \left( 0\right) \right\}
\right\rangle -\left\langle \Phi \right\rangle \left\langle \Psi
\right\rangle \right]  \label{CSI-gS} \\
\xi ^{Sg}(\vec{r}) &=&\int_{0}^{\infty }d\tau \left[ \frac{1}{2}\left\langle
\left\{ \Psi \left( \tau \right) ,\Phi \left( 0\right) \right\}
\right\rangle -\left\langle \Psi \right\rangle \left\langle \Phi
\right\rangle \right]  \label{CSI-Sg} \\
\xi^{SS} (\vec{r}) & = & \int_{0}^{\infty}d\tau\left[ \frac{1}{2}%
\left\langle \left\{ \Psi\left( \tau\right) ,\Psi\left( 0\right) \right\}
\right\rangle -\left\langle \Psi\right\rangle ^{2}\right].  \label{CSI-SS}
\end{eqnarray}
The cross diffusion term proportional to $\eta_{ij}(\vec{r})$ is usually
much smaller than the diffusion terms due to the dipole force fluctuations $%
D_{ij}(\vec{r})$ and due to spontaneous emission $\propto E_{ij}$ \cite
{doherty,cohen}, and we shall neglect it. Using this approximation and the
above expression for the diffusion matrix, the six-dimensional phase space
diffusion matrix of the Fokker-Planck equation (\ref{Fokker-P}) becomes
semipositive-definite. This means that it can be associated to a classical
phase space stochastic process, describing the stochastic trajectories of
the atomic center-of-mass within the cavity. Consistently with the adiabatic
and quasi-classical description discussed above, the motional Wigner
function $f(\vec{r},\vec{p},t)$ becomes therefore a nonnegative, classical
phase space probability distribution \cite{cohen}. Our numerical analysis is
based just on the simulation of these stochastic 3-D trajectories, which are
obtained as solutions of the It\^{o} stochastic equations associated to the
Fokker-Planck equation (\ref{Fokker-P}). Moreover, as it can be easily seen
from Eq.~(\ref{modo-cavita}), (\ref{ciambella}), and Eqs.~(\ref{fr-gg})-(\ref
{diff-SS}), one has $\Gamma_{xx} \propto \lambda_{g}^{-2}\sim
\lambda_{S}^{-2}$, $\Gamma_{xy} \sim \Gamma_{xz}\propto
\lambda_{g}^{-1}W_{g}^{-1}\sim \lambda_{S}^{-1}W_{S}^{-1}$, $\Gamma_{yy}\sim
\Gamma_{zz}\propto W_{S}^{-2}$, and the same is true for the diffusion
matrix $D_{ij}$. Since it is always $\lambda_{g},\lambda_{S} \ll W_{g},W_{S}$%
, it is evident that the only relevant term in the friction and diffusion
matrices is the $xx$ component along the cavity axis, where both the
quantized field and the FORT mode show the largest spatial gradients \cite
{vanenk}. In our numerical simulations, we have therefore considered both
the friction force and the dipole force contribution to diffusion, only
along the cavity axis $x$, while we have kept the spontaneous emission
diffusion terms in all three directions.

Taking into account Eq.~(\ref{Fokker-P}), and the above approximations, we
have therefore numerically solved the following It\^{o}-equations \cite
{Risken,gardiner2}: 
\begin{eqnarray}
d\vec{r} & = &\frac{\vec{p}}{M}dt  \nonumber \\
d\vec{p} & =& \vec{\phi}(\vec{r}) dt-\left( \matrix{ \Gamma_{xx}(\vec{r}) &
0 & 0 \cr 0 & 0 & 0 \cr 0 & 0 & 0 \cr} \right) \left( \matrix{ p_{x}\cr
p_{y} \cr p_{z} \cr} \right) dt+\left( \matrix{
\sqrt{D_{xx}(\vec{r})}dW_{1}\cr 0 \cr 0 \cr} \right)  \label{ITO} \\
& +& \hbar k_{a}\sqrt{\gamma\left\langle \sigma^{\dagger}\sigma\right\rangle
(\vec{r})}\left( \matrix{ \sqrt{E_{xx}}dW_{x}\cr \sqrt{E_{yy}}dW_{y} \cr
\sqrt{E_{zz}}dW_{z} \cr} \right)  \nonumber
\end{eqnarray}
where $dW_{1}$ and $d\vec{W}\equiv(dW_{x},dW_{y},dW_{z})$ are four
independent, zero mean, Wiener increments with the property $%
dW_{i}dW_{j}=2\delta_{ij}dt$, $(i,j=1,x,y,z)$. The quantities $\vec{\phi}(%
\vec{r})$, $\Gamma_{xx}(\vec{r})$, $D_{xx}(\vec{r})$, $\left\langle
\sigma^{\dagger}\sigma\right\rangle (\vec{r})$ in Eq.~(\ref{ITO}) have been
determined by numerically solving Eq.~(\ref{LBloch}) in the atom-cavity
Hilbert space truncated at $n=4$ photons (see \cite{Tan} for details). We
have also checked that these numerical solutions reproduce the results of
the analytical approach of \cite{ritsch} in the weak-driving limit.

\section{Numerical results}

In the next subsections we present and discuss the results of the numerical
simulations in both cases, (a) and (b). We have chosen equivalent conditions
for the two cases, so that the corresponding numerical results are directly
comparable.

\subsection{Study of case (a)}

We consider parameter values referred to the experiment of Ref.~\cite{Ye-exp}%
. In fact, we consider the Cs transition at $\lambda_{a}=2\pi/k_a = 852.4$
nm between the ground state $|g \rangle =| 6S_{1/2},F=4,m_{F}=4 \rangle $
and the excited state $|e \rangle =| 6P_{3/2},F=5,m_{F}=5\rangle $, and a
cavity mode resonant with it, i.e., $\omega_{ap}=\omega_{gp}=-\Delta_{p} =
2\pi\times10$ MHz ($\Delta_{p}$ is the probe detuning from resonance). The
spontaneous emission rate is $\gamma = 2\pi \times 2.6$ MHz, while the other
quantized mode parameters are $g_{0}=2\pi \times 30 \sqrt{e}$ MHz, $\kappa =
2\pi \times 4$ MHz, $W_{g}= 20$ $\mu$m, and $E=6.77$ MHz, so that the empty
cavity mean photon number is $N_{e}=E^{2}/\left(\kappa^{2}+\Delta_{p}^{2}%
\right)=0.01$. Since it is $g_{0}>\gamma,\kappa$, we are therefore in the
strong coupling regime of cavity QED. Then we use a LG$_{01}$ doughnut mode
as red-detuned FORT field, with parameter values $W_{S}=20$ $\mu$m, $%
S_{0}=\pi e/2$ MHz/$\mu {\rm m}^{2}$. These choices give $\rho_{\max}\sim
14.1$ $\mu$m for the doughnut radius, and the maximum Stark shift given by
the FORT mode, achieved at $\rho_{\max}$, is $S_{\max}=2\pi\times50$ MHz.

As discussed above and in \cite{vanenk}, we have chosen a fictitious large
value of $\lambda_{S}$, so that $L= 15\lambda_{S}=16\lambda_{g}$ ($\lambda_g
= \lambda_{a}$) in order to simplify the simulation, without however
changing the physics because the exact value of $\lambda_{S}$ is unimportant
as long as it is far-off resonance. The dominant potential is the one due to
the FORT and therefore the atomic equilibrium positions will be situated at
the FORT antinodes $x_{n}^{e}=(n-1/2)\lambda_{S}/2$, $n=1,\ldots \ldots,$
along the axial direction and at the nonzero radial distance from the cavity
axis $\rho_{\max}\sim 14.1$ $\mu$m. Because of the sharp radial potential
well of the doughnut mode, the atom will experience a sharp radial
confinement around $\rho_{\max}$ besides the axial confinement around an
antinode.

\begin{figure}[h]
\centerline{\epsfig{figure=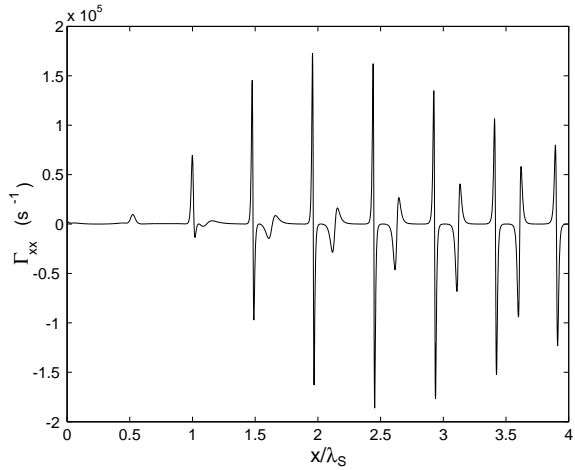,width=3.0in}}
\caption{ Axial friction coefficient $\Gamma_{xx}(\vec{r})$, at fixed radial
distance $\protect\rho=\protect\rho_{\max}$ (bottom of the radial well),
versus the rescaled axial coordinate $x/\protect\lambda_{S}$, in the case
(a) of opposite Stark shifts induced by the FORT mode. Parameter values are
in the text. }
\label{dampa}
\end{figure}

\begin{figure}[h]
\centerline{\epsfig{figure=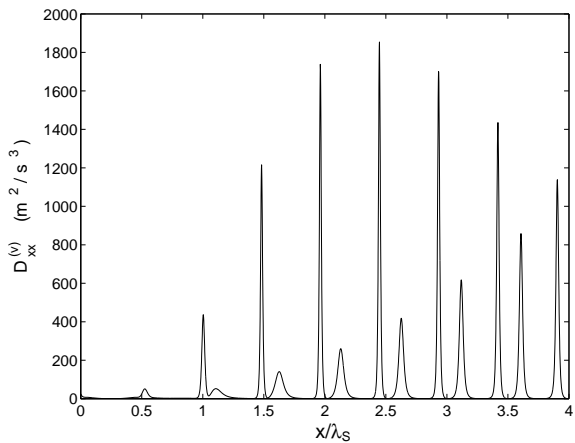,width=3.0in}}
\caption{ Dipole contribution to the axial velocity diffusion coefficient $%
D_{xx}(\vec{r})/M^{2}$, at fixed radial distance $\protect\rho=\protect\rho%
_{\max}$ (bottom of the radial well), versus the rescaled axial coordinate $%
x/\protect\lambda_{S}$, in the case (a) of opposite Stark shifts induced by
the FORT mode. Parameter values are in the text. }
\label{diffa}
\end{figure}

The superposition of the two modes with different wavelengths determines a
spatially aperiodic situation within the cavity. The above choice gives $8$
not equivalent potential wells in which the atom is subject to different
couplings with the quantized cavity mode. This aperiodic situation is shown
in Figs.~\ref{dampa} and \ref{diffa}, where the axial friction coefficient $%
\Gamma_{xx}(\vec{r})$ of Eqs.~(\ref{friction}) and (\ref{fr-gg})-(\ref{fr-SS}%
), and the dipole contribution to the axial diffusion coefficient $D_{xx}(%
\vec{r})$ of Eqs.~(\ref{diffusione}) and (\ref{diff-gg})-(\ref{diff-SS}), at
fixed radial distance $\rho=\rho_{\max}$ (bottom of the radial well), and as
a function of the axial coordinate $x$, are plotted.

\subsubsection{Simulation results\label{cond_media}}

To characterize the trapping and cooling dynamics, we have carried out a
quantitative study inside a central well, the one centered in $x_{5}
^{e}=2.25\lambda_{S}$ and ranging from $2\lambda_{S}$ to $2.5\lambda_{S}$. 
In order to simulate the
typical experimental condition, we have chosen proper initial conditions,
considering that, in the experimental procedure, the FORT field is switched
on only when the laser probe transmission exceeds a fixed threshold
indicating the presence of the atom inside the cavity. The initial position
has been taken axially $\lambda_{S}/8$ away from equilibrium point $%
x_{5}^{e} $ (then $x_{0}=2.125\lambda_{S}$), radially along the doughnut
maximum $\rho_{0}=\rho_{\max}$, and uniformly distributed over the polar
angle $\theta $. For what concerns the initial velocity, it is reasonable to
choose a vertical velocity with components $v_{x0}=v_{y0}=0$, $v_{z0}=10$
cm/s.

\begin{figure}[h]
\centerline{\epsfig{figure=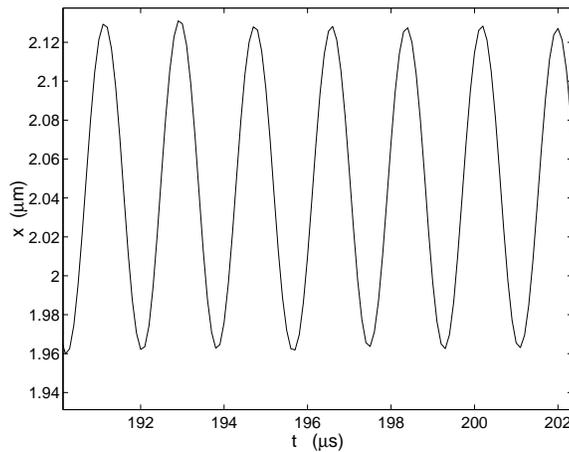,width=3.0in}}
\caption{ Time evolution of the axial position in the case of an initial
velocity tangential with respect to the doughnut FORT mode. Parameter values
are in the text. }
\label{axtraj}
\end{figure}

\begin{figure}[h]
\centerline{\epsfig{figure=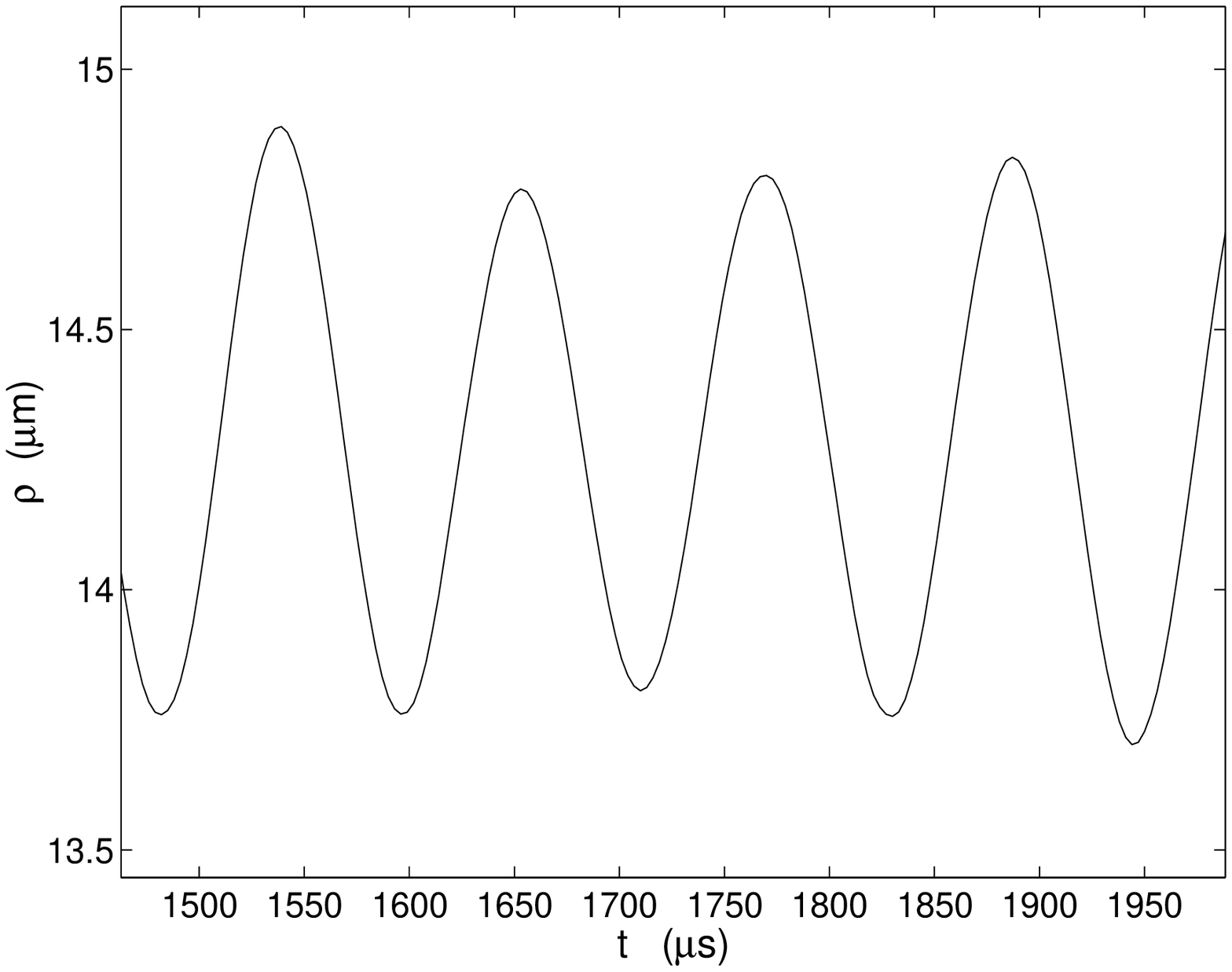,width=3.0in}}
\caption{ Time evolution of the radial coordinate in the case of an initial
velocity tangential with respect to the doughnut FORT mode. Parameter values
are in the text. }
\label{radtraj}
\end{figure}

\begin{figure}[h]
\centerline{\epsfig{figure=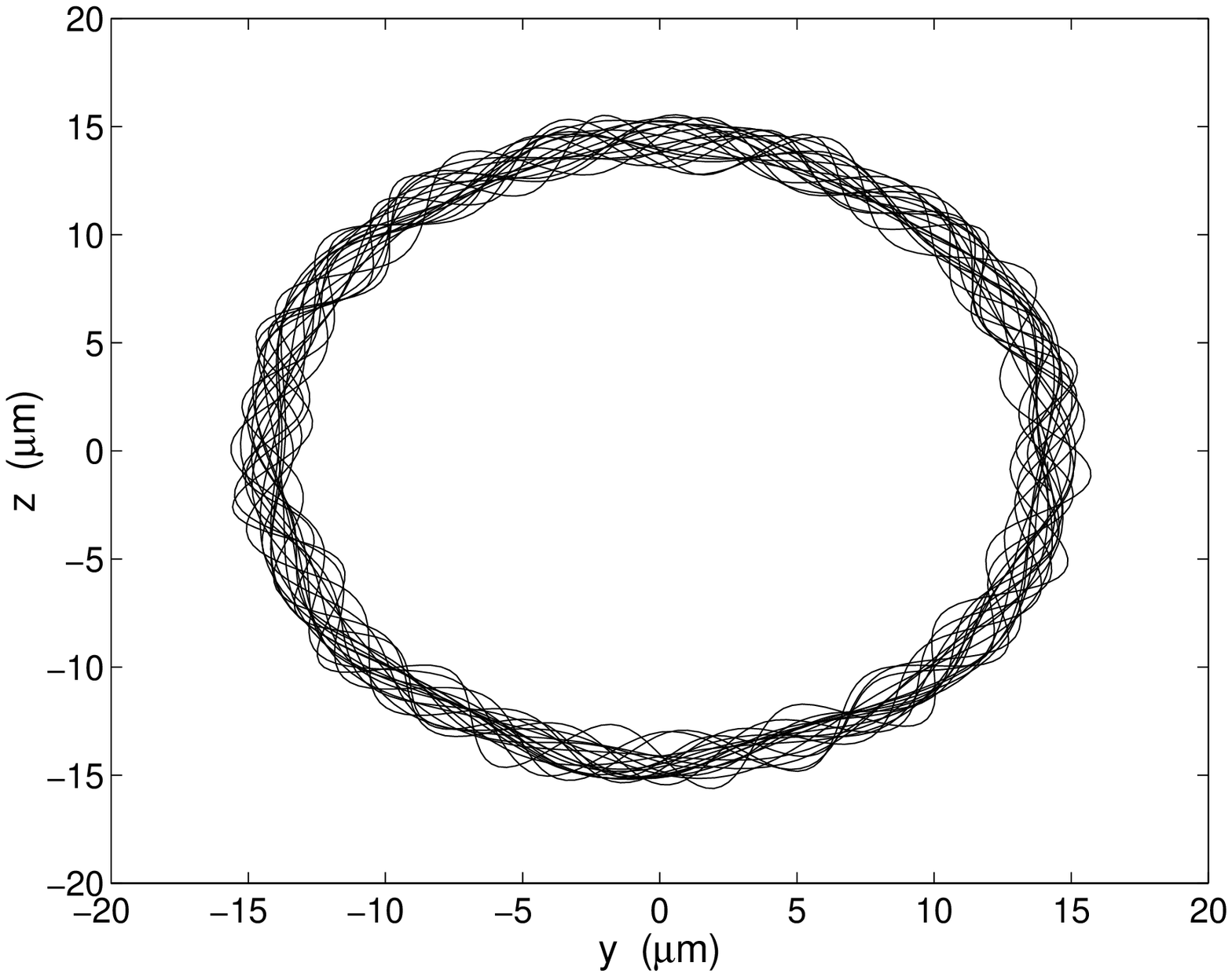,width=3.0in}}
\caption{ Radial trajectory in the case of an initial velocity tangential
with respect to the doughnut FORT mode. Parameter values are in the text. }
\label{radtraj2}
\end{figure}

\begin{figure}[h]
\centerline{\epsfig{figure=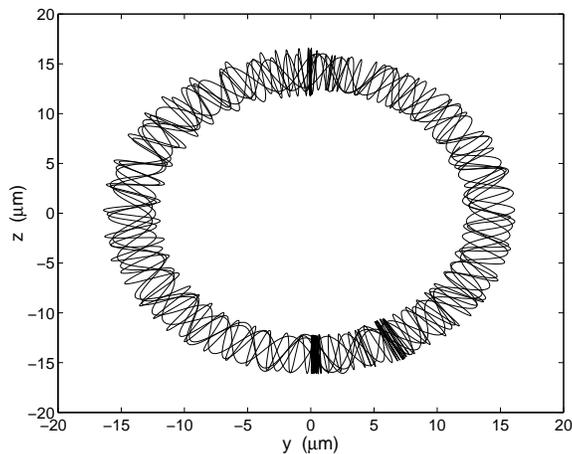,width=3.0in}}
\caption{ Radial trajectory in the case of an initial velocity orthogonal
with respect to the doughnut FORT mode. Parameter values are in the text. }
\label{radtraj3}
\end{figure}

In order to examine qualitatively the typical atomic motion, we report some
snapshots from two simulated trajectories with slightly different initial
conditions: one describes a tangential incidence of the atom along the
doughnut perimeter ($x_{0}=2.125\lambda _{S}$, $y_{0}=\rho _{\max }$, $%
z_{0}=0$, $v_{x0}=v_{y0}=0$, $v_{z0}=10$ cm/s) (see Figs.~\ref{axtraj}-\ref
{radtraj2}); the other instead describes an orthogonal incidence with
respect to the doughnut mode ($x_{0}=2.125\lambda _{S}$, $y_{0}=0$, $%
z_{0}=\rho _{\max }$, $v_{x0}=v_{y0}=0$, $v_{z0}=10$ cm/s) (see Fig.~\ref
{radtraj3}).

The trajectory is well confined, axially around a FORT antinode, and
radially around the maximum $\rho _{\max }$: the 3D atomic motion occurs
substantially on a plane orthogonal to the cavity axis, with the trajectory
drawing just the shape of the doughnut mode. From Figs.~\ref{axtraj}-\ref
{radtraj3} one can recognize that the atomic motion is characterized by
three different time scales:

\begin{enumerate}
\item  {The fastest timescale is given by the axial oscillations, which
have, for our parameter values, a time period $\sim 2\mu $s (see Fig.~\ref
{axtraj}).}

\item  {A slower timescale is associated with the radial oscillations,
characterized by a time period $\sim 100$ $\mu $s (see Fig.~\ref{radtraj});
these oscillations become wider in the case of orthogonal incidence with
respect to the doughnut mode, since, in this case, the atom probes more the
doughnut radial elasticity (compare, in fact, Fig.~\ref{radtraj2} where the
oscillation amplitude is $\sim 1$ $\mu $m with Fig.~\ref{radtraj3}, where
the amplitude is $\sim 4$ $\mu $m).}

\item  {The third and slowest timescale is given by the atomic rotations
around the cavity axis. For tangential incidence, the initial angular
momentum is large and the rotation period is $\sim 1$ ms while, for
orthogonal incidence, the atom acquires a nonzero angular momentum only
because of radial diffusion, and the rotation period is larger, $\sim 10$ ms.%
}
\end{enumerate}

The final escape of the atom from the cavity is practically always along the
axial direction, and this is due to the heating provided by the strong axial
diffusion, which prevails with respect to the radial diffusion, determined
only by the spontaneous emission.

\begin{figure}[h]
\centerline{\epsfig{figure=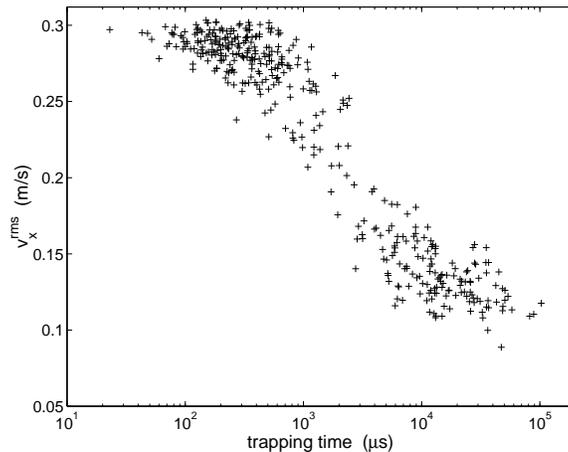,width=3.0in}}
\caption{ Root mean square velocity along the cavity axis $v_{x}^{rms}$ as a
function of the trapping time $T$, for each simulated trajectory, in the
case (a) of opposite Stark shifts. }
\label{vTa}
\end{figure}

\begin{figure}[h]
\centerline{\epsfig{figure=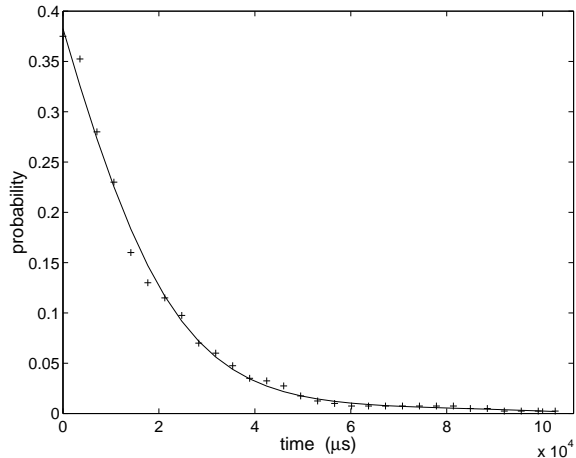,width=3.0in}}
\caption{ Trapping time statistics for the subset of trapped atoms in the
case (a) of opposite Stark shifts. $P(t)$ is the probability for an atom to
be trapped for a time greater than $t$. The full line is a fitting decay
curve yielding a mean trapping time $\protect\tau =(17\pm1)$ ms. }
\label{pTa}
\end{figure}

To determine the mean trapping times, we have defined as trapping time $T$
the time spent by the atom inside a single potential well, $\lambda_{S}/2$
wide along the axial direction, and with a radius equal to $2W_{S}$.
Sampling about 400 simulated trajectories, we have found the results
displayed in Figs.~\ref{vTa} and \ref{pTa}.

In Fig.~\ref{vTa} we have displayed the root mean square velocity along the
cavity axis $v_{x}^{rms}$ as a function of the trapping time $T$ for each
simulated trajectory. We can see a clear separation of simulated points. In
fact, about $60\%$ of atoms are not trapped at all: they correspond to the
upper set of points in Fig.~\ref{vTa}, having a velocity $v_{x}^{rms}> 20$
cm/s, and for which the trapping time is below $2$ ms. These are not cooled
via the cavity QED interaction, and the velocity $v_{x}^{rms}\sim 28$ cm/s
is mainly a result due to the initial conditions. These uncooled atoms are
those more influenced by the spatial region where the axial friction is
negative (see Fig.~\ref{dampa}), where atoms can be accelerated. The
remaining $40\%$ of atoms (the points below the threshold of $20$ cm/s in
Fig.~\ref{vTa}) are trapped: their velocity $v_{x}^{rms}$ of about $14$ cm/s
is a result of the cooling provided by the exchange of cavity photons. These
atoms reach thermal equilibrium and have trapping times greater than $2$ ms.

Considering only the subset of trapped atoms, they have a probability $P(t)$
to be trapped for a time greater than $t$. The trapping time statistics is
shown in Fig.~\ref{pTa}, where it is also compared with a decaying fitting
curve $P^{th}$ (full line in Fig.~\ref{pTa}). The best fitting mean trapping
time is $\tau =\int_{0}^{\infty}t\left( - dP^{th}/dt\right) dt=(17\pm1)$ ms,
which is comparable to the experimental value obtained using a fundamental
Gaussian FORT mode in Ref.~\cite{Ye-exp} and with the numerical simulations
performed in \cite{vanenk}, again for a fundamental Gaussian FORT mode. 
This is not
surprising because, except for the fact that the atom is now trapped at a
nonzero distance from the cavity axis, the physics of cooling is similar to
that occurring in a lowest-order Gaussian FORT mode, 
and the analysis of Ref.~\cite{vanenk} can be essentially repeated.

\subsection{Study of case (b)}

We consider the same parameter values of case (a), except that we slightly
adapt the probe detuning and choose a value $\Delta_{p}=-2\pi\times35$ MHz.
As a consequence, we then set $E=22.13$ MHz, in order to keep the same
empty-cavity mean photon number $N_{e}=0.01$ of case (a).

\begin{figure}[h]
\centerline{\epsfig{figure=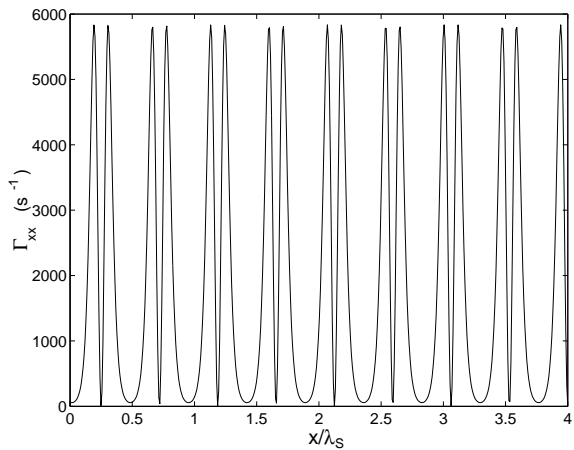,width=3.0in}}
\caption{ Axial friction coefficient $\Gamma_{xx}(\vec{r})$, at fixed radial
distance $\protect\rho=\protect\rho_{\max}$ (bottom of the radial well),
versus the rescaled axial coordinate $x/\protect\lambda_{S}$, in the case
(b) of equal Stark shifts induced by the FORT mode. }
\label{dampb}
\end{figure}

\begin{figure}[h]
\centerline{\epsfig{figure=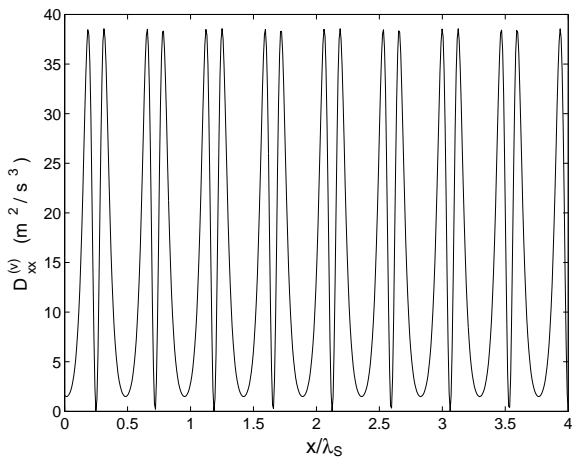,width=3.0in}}
\caption{ Dipole contribution to the axial velocity diffusion coefficient $%
D_{xx}(\vec{r})/M^{2}$, at fixed radial distance $\protect\rho=\protect\rho%
_{\max}$ (bottom of the radial well), versus the rescaled axial coordinate $%
x/\protect\lambda_{S}$, in the case (b) of equal Stark shifts induced by the
FORT mode. }
\label{diffb}
\end{figure}

The situation is in many respects very similar to the preceding one: the
atomic equilibrium positions are again situated at the FORT antinodes $%
x_{n}^{e}=(n-1/2)\lambda_{S}/2$, $n=1,...$ along the axial direction and at
the nonzero radial distance from the cavity axis $\rho_{\max}\sim 14.1$ $\mu$%
m. There is still an aperiodic situation with $8$ not equivalent wells within
the cavity. However, for equal atomic shifts, the FORT does not affect both
friction and diffusion (it does not affect the force fluctuations, see 
Eqs.~(\ref{friction}) and (\ref{diffusione})) and therefore, for what concern
friction and diffusion, an axially periodic situation is restored in this
case, with a spatial period set by the cavity QED mode. The periodic
friction and diffusion spatial variations are shown in Figs.~\ref{dampb} and 
\ref{diffb}, where the axial friction coefficient $\Gamma_{xx}(\vec{r})$ of
Eqs.~(\ref{friction}) and (\ref{fr-gg})-(\ref{fr-SS}), and the dipole
contribution to the axial diffusion coefficient $D_{xx}(\vec{r})$ of Eqs.~(%
\ref{diffusione}) and (\ref{diff-gg})-(\ref{diff-SS}), at fixed radial
distance $\rho=\rho_{\max}$, and as a function of the axial coordinate $x$,
are plotted.

If we compare Figs.~\ref{dampa} and \ref{diffa} with Figs.~\ref{dampb} and 
\ref{diffb} we see that in case (b) the maxima of the axial friction and
diffusion coefficients are lower, but one has the advantage that now the
friction coefficient is always positive, while in case (a) it may assume
very large negative values, yielding heating rather than cooling. This
always positive friction implies that all atoms falling in the cavity are
now cooled and trapped, while in case (a), a fraction of the atoms can be
heated and are not trapped.

\subsubsection{Simulation results}

\begin{figure}[h]
\centerline{\epsfig{figure=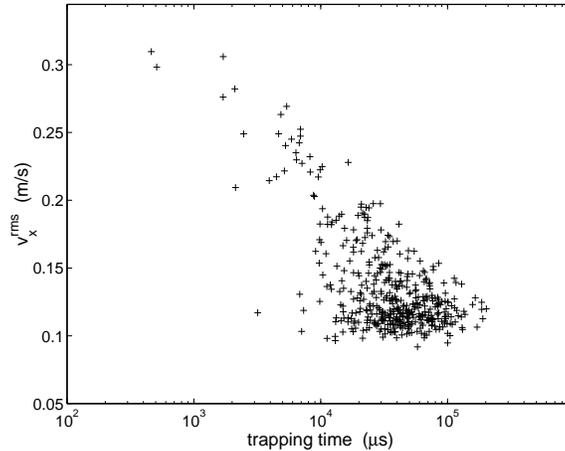,width=3.0in}}
\caption{ Root mean square velocity along the cavity axis $v_{x}^{rms}$ as a
function of trapping time $T$, for each simulated trajectory, in the case
(b) of equal Stark shifts. }
\label{vTb}
\end{figure}

\begin{figure}[h]
\centerline{\epsfig{figure=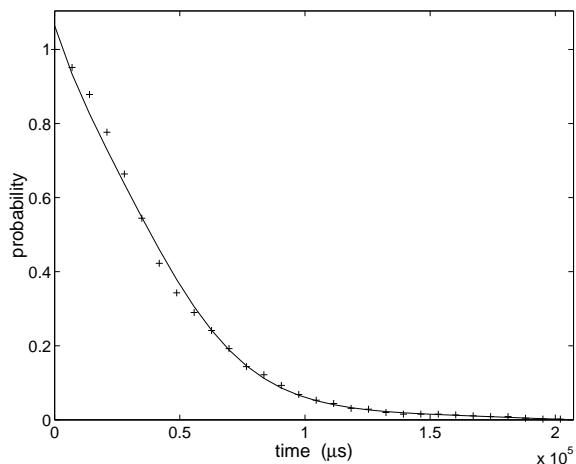,width=3.0in}}
\caption{ Probability $P(t)$ for an atom to be trapped for a time greater
than $t$ in the case (b) of equal Stark shifts. The full line is a fitting
decay curve, yielding a mean trapping time $\protect\tau =(44\pm 3)$ ms. }
\label{pTb}
\end{figure}

In order to calculate the mean trapping time in case (b), we have carried
out a computer simulation considering the same initial condition of case (a)
(see section \ref{cond_media}). Sampling about 400 simulated trajectories,
we have found the results shown in Figs.~\ref{vTb} and \ref{pTb}.

In Fig.~\ref{vTb} we have displayed the root mean square velocity along the
cavity axis $v_{x}^{rms}$ as a function of trapping time $T$ for each
simulated trajectory. At variance with case (a), now all atoms are trapped,
with trapping times greater than $1$ ms. This is due the fact that axial
friction is always positive and atoms are cooled everywhere. In this case
the probability $P(t)$ to be trapped for a time greater than $t$ (shown in
Fig.~\ref{pTb}) is computed considering all simulated points. The data are
again fitted by a decaying fitting function $P^{th}(t)$ (full line in Fig.~%
\ref{pTb}), yielding a mean lifetime $\tau =\int_{0}^{\infty}t\left( -
dP^{th}/dt\right) dt=(44\pm 3)$ ms.

This is the most relevant result of our investigation, showing that using a
doughnut mode as FORT mode allows to achieve significant trapping times in
the case (b), when the FORT induces equal Stark shifts on the atomic levels.
In fact, we get a mean trapping time larger than that obtained, in the same
situation, with a red-detuned fundamental Gaussian mode 
(see the numerical analysis of
Ref.~\cite{vanenk}, Section IVF, where $\tau =(28\pm 2)$ ms). This shows
that, in the case of equal Stark shifts, the radial confinement at a nonzero
distance from the cavity axis provided by the doughnut FORT mode is
preferable with respect to the radial trapping along the cavity axis
provided by the TEM$_{0,0}$ FORT mode. The improvement provided by doughnut
mode is useful for quantum information processing applications in cavity QED
systems, because when the FORT mode induces equal shifts on the ground and
excited state, the two states are trapped at the same positions within the
cavity, and the internal state can be manipulated independently of its
center-of-mass state. In quantum information applications, it is important
to evaluate the variation of the cavity-QED coupling $\Delta g$. 
In our parameters' regime it is $\Delta g\succsim 16\%$, 
which is worser than that for a TEM$_{0,0}$
FORT mode under the same conditions ($\Delta g\succsim 6\%$). However the
problem associated with such a variation of the coupling can be circumvented
if one adopts an adiabatic transfer scheme such as the one
suggested in \cite{hotatoms} for quantum information
processing with hot trapped atoms. In such a case, 
the operation time $T_{op}$
for the adiabatic transformations must satisfy the conditions $2\mu s\sim 
\nu _{axial}^{-1} < T_{op} \ll \nu _{radial}^{-1}
\sim 100\mu s$, where $\nu_{axial,radial}$ 
are the axial or radial oscillation frequencies.

An interesting alternative to the configuration studied here
is to replace
the TEM$_{0,0}$ cavity QED mode with the higher order
LG$_{0,1}$ cavity QED mode, having the
same coupling $g$ at $\rho _{\max }$.
In this way we achieve a better overlap between the FORT mode and cavity
QED one, which leads to a smaller variation of coupling
$\Delta_g$, and therefore also to a better cooling configuration.
In fact, repeating the simulations with the LG$_{0,1}$ cavity QED mode
we have seen a slight improvement of the trapping time.

It is clear that the additional practical difficulty of using another LG mode
makes the experimental realization even harder; however this should be very
challenging since, extending the configuration from one LG cavity-QED mode
to many degenerate LG cavity-QED modes, one has new
possibilities, as following atomic motion in detail \cite{kaleidoscopio}, or
increasing the cooling effect as shown in \cite{Scaling}.

\section{Discussion of the results}

\begin{figure}[h]
\centerline{\epsfig{figure=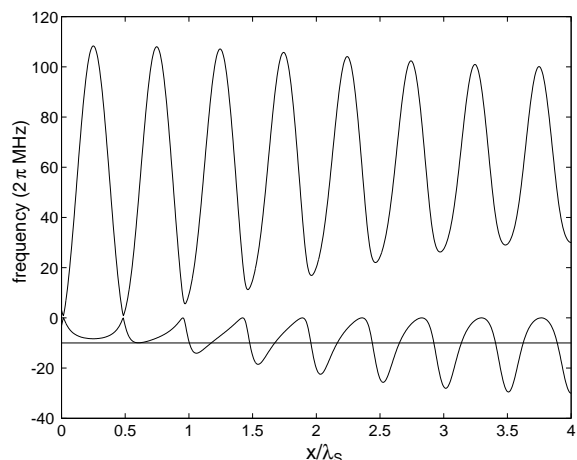,width=3.0in}}
\caption{ Transition frequencies $\Delta_{\pm}$ (relative to $\protect\omega%
_{a}$) from the ground state $\left| 0\right\rangle $ to the first excited
dressed states $\left| \pm\right\rangle $ along the axial direction at fixed
radial distance $\protect\rho=\protect\rho_{\max}$ (bottom of the radial
well), in the case (a) of opposite Stark shifts. The straight line gives the
probe detuning $\Delta_{p}$. }
\label{transia}
\end{figure}

\begin{figure}[h]
\centerline{\epsfig{figure=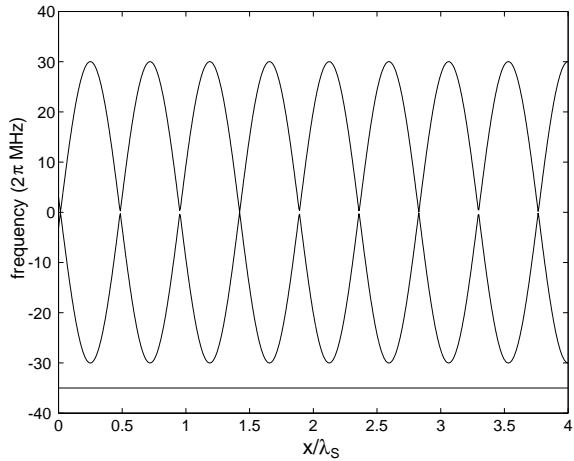,width=3.0in}}
\caption{ Transition frequencies $\Delta_{\pm}$ (relative to $\protect\omega%
_{a}$) from the ground state $\left| 0\right\rangle $ to the first excited
dressed states $\left| \pm\right\rangle $ along the axial direction at fixed
radial distance $\protect\rho=\protect\rho_{\max}$ (bottom of the radial
well), in the case (b) of equal Stark shifts. The straight line gives the
probe detuning $\Delta_{p}$. }
\label{transib}
\end{figure}

Let us now discuss in detail which are the main features of using
red-detuned doughnut modes as FORT fields. The atom-cavity dressed picture
provides an intuitive way to understand the advantages brought by the use of
the doughnut mode. In the present case of very weak driving ($N_{e} = 0.01$
in our case), the cooling mechanism is well described in terms of the
eigenstates of the atom-cavity system (the dressed states) containing at
most one excitation (see also \cite{ritsch,vanenk,Vernooy}). The state with
no excitation is the ground state $|0\rangle$, with energy 
$E_{0} = -\hbar S(
\vec{r})$, while the first two dressed states with one excitation $%
|\pm\rangle $ have energies 
\begin{equation}
E_{\pm} = \left\{\matrix
{\hbar\omega_{a}\pm\hbar\sqrt{g^{2}(\vec{r})+S^{2}(\vec{r})} & {\rm case
(a)} \cr \hbar\omega_{a}-\hbar S(\vec{r})\pm\hbar g(\vec{r}) & {\rm case
(b)} \cr} \right.,  \label{livelli-dressed}
\end{equation}
so that the transition frequencies (relative to $\omega_{a}$) from the
ground state $\left| 0\right\rangle $ to excited states $\left|
\pm\right\rangle $ have the expressions 
\begin{equation}
\Delta_{\pm}\equiv\frac{(E_{\pm}-E_{0})}{\hbar}-\omega_{a}= \left\{%
\matrix{S(\vec{r} )\pm\sqrt{g^{2}(\vec{r})+S^{2}(\vec{r})} & {\rm case (a)}
\cr \pm g(\vec{r}) & {\rm case (b)} \cr} \right..  \label{transition-freq}
\end{equation}
The spatial variation of these transition frequencies along the axial
direction (and at the radial position $\rho = \rho_{max}$) is shown in Fig.~%
\ref{transia} for the case (a) of opposite Stark shifts, and in Fig.~\ref
{transib} for the case (b) of equal Stark shifts.

The patterns of spatial variation of the transition frequency allow to
understand the difference between cases (a) and (b). The cooling mechanism
provided by the cavity QED interaction in the weak driving limit, can be
understood in analogy with Doppler cooling. In fact, by tuning below
resonance, stimulated absorption of a probe photon followed by spontaneous
emission or cavity decay leads to a loss of energy. The maximum cooling rate
is achieved when the excitation rate times the detuning is maximum. If the
detuning turns from red to blue, cooling is replaced by heating, while if
the detuning becomes more red, the atom is still cooled, but at a lower
rate. In case (a) (Fig.~\ref{transia}), it is better to tune the probe to
the lower dressed state because it has smaller spatial variations (see the
straight line in Fig.~\ref{transia}) and it is easier to reach a compromise
between having cavity regions with an optimal cooling rate (red detuning)
and not too large regions with heating (blue detuning). However, as the atom
moves radially from the doughnut intensity maximum at $\rho = \rho_{max}$,
the situation rapidly worsens, either if the atom moves towards the center
(the FORT beam decreases while the quantized field increase and the blue
detuned region increases) or if tends to leave the cavity (the detuning
becomes more red and the cooling rate decreases).

In case (b), one tunes again to the lowest dressed state in order to have
red detuning, and therefore cooling, throughout the cavity, when the atom is
at the radial equilibrium position $\rho_{max}$ (see the straight line in
Fig.~\ref{transib}). From Eq.~(\ref{transition-freq}) we see that the probe
detuning decreases (becomes less red) when the atom moves radially towards
the center, while it becomes more red if the atom moves away radially.
However, the probe detuning can be chosen so to remain always red: in this
way the atom is cooled everywhere, and is never accelerated. For this reason
the doughnut FORT mode provides longer trapping times in case (b) of equal
Stark shifts rather than in case (a) of opposite shifts. What is more
important is that, in case (b), the doughnut FORT mode provides longer
trapping times than a fundamental Gaussian FORT mode (see \cite{vanenk},
Section IVF). In fact, in the latter case, the probe detuning would be tuned
so to have optimal cooling on the cavity axis, where the atoms will be now
trapped. However, because of radial diffusion (due to spontaneous emission)
which also leads to an increasing angular momentum and the consequent rising
of a centrifugal potential, the atom tends to move radially away from the
cavity axis, so that the optimal cooling condition is rapidly lost. In fact,
for increasing $\rho $, the axial pattern of Fig.~\ref{transib} rapidly
vanishes, the driving probe becomes too far-off resonance and the cooling
efficiency is lost.

The advantage of the doughnut FORT with respect to the TEM$_{0,0}$ FORT is that
it imposes a much sharper radial well around the equilibrium position
corresponding to the optimum cooling condition. The atom is less free to
move radially, and moreover is subject to a smaller centrifugal force
because it is trapped by the doughnut at a larger radial distance from the
cavity axis. In other words, the radial potential well provided by the
doughnut FORT mode is more suitable to counteract the radial departure of
the atom and so to preserve the optimal cooling condition.

\begin{figure}[h]
\centerline{\epsfig{figure=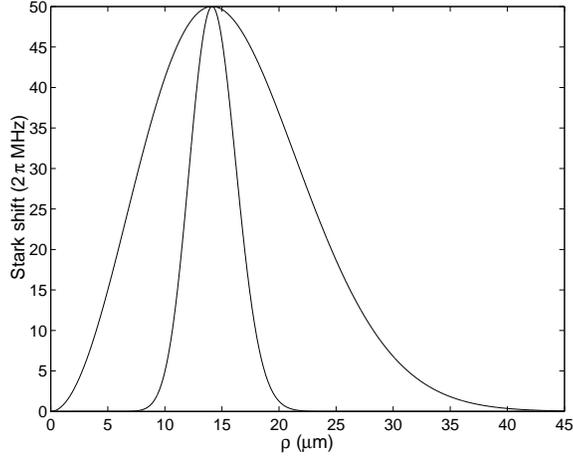,width=3.0in}}
\caption{ Stark shifts $S_{0,1}$ and $S_{0,12}$ (narrower curve) as a
function of the radial coordinate $\protect\rho$, at the axial antinode.
Parameter values are in the text.}
\label{compa1-12}
\end{figure}

The importance of a strong radial confinement to optimize cooling and
trapping in the case of equal Stark shifts can be illustrated also checking
that when the width of the radial well is decreased, the trapping time
increases. To this purpose, we have carried out a simulation where a higher
order Laguerre-Gauss FORT mode, LG$_{0,12}$, is used instead of the LG$_{0,1}
$ FORT mode, keeping the other conditions unchanged. To be more specific, we
have considered the same experimental parameters of the case with the LG$%
_{0,1}$ FORT mode in case (b), in such a way that the LG$_{0,12}$ mode gives
the same Stark shift $S_{\max }=2\pi \times 50$ MHz at the same radial
distance $\rho _{\max }\sim 14.1$ $\mu m$. This means adapting both the
incident FORT power and the doughnut mode waist, so that the only difference
with the case studied in the preceding Section is the smaller width of the
radial well (see Fig.~\ref{compa1-12}, where the radial profile of the Stark
shift induced by the LG$_{0,12}$ FORT mode is compared with that of the LG$%
_{0,1}$ mode). The chosen parameters are the same as above except that now
the spatial dependence of the Stark shift (equal for both ground and excited
state) is 
\begin{equation}
S(\rho ,x)=S_{0}\rho ^{24}\sin ^{2}(k_{S}x)\exp \left\{ \frac{-2\rho ^{2}}{%
W_{S}^{2}}\right\} ,  \label{ciamb12}
\end{equation}
with $W_{S}=20/\sqrt{12}\sim 5.77$ $\mu $m and $S_{0}\sim 1.2\times 10^{-20}$
MHz$/\mu $m$^{24}$.

\begin{figure}[h]
\centerline{\epsfig{figure=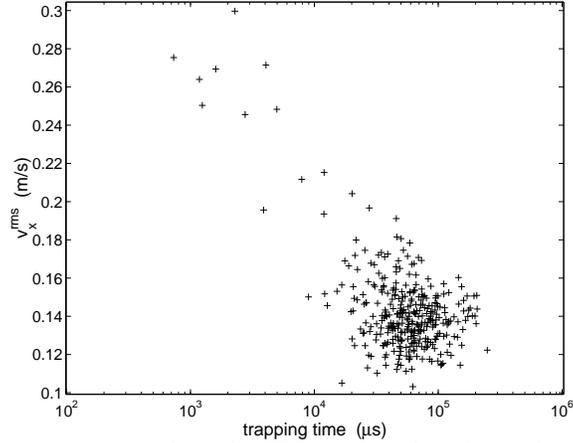,width=3.0in}}
\caption{ Root mean square velocity along the cavity axis $v_{x}^{rms}$ as a
function of trapping time $T$, for each simulated trajectory in the case (b)
and with the higher order FORT mode LG$_{0,12}$. }
\label{vTb12}
\end{figure}

\begin{figure}[h]
\centerline{\epsfig{figure=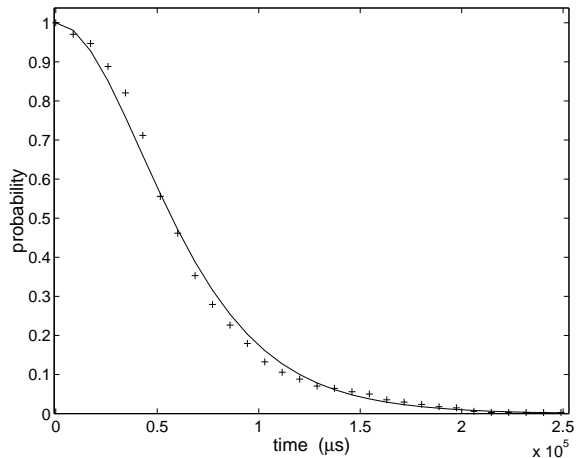,width=3.0in}}
\caption{ Probability $P(t)$ for an atom to be trapped for a time greater
than $t$ in the case (b) and with the higher order FORT mode LG$_{0,12}$.
The full line is a fitting decay curve yielding a mean trapping time $%
\protect\tau = (64 \pm 5)$ ms. }
\label{pTb12}
\end{figure}

Considering again the initial conditions specified in Section \ref
{cond_media} and sampling about 400 simulated trajectories, we have found
the results shown in Figs.~\ref{vTb12} and \ref{pTb12}.

In Fig.~\ref{vTb12} we have displayed the root mean square velocity along
the cavity axis $v_{x}^{rms}$ as a function of trapping time $T$ for each
simulated trajectory. All atoms are again trapped, with trapping times
greater than $1$ ms. The probability $P(t)$ to be trapped for a time greater
than $t$ (shown in Fig.~\ref{pTb12}) is computed considering all simulated
points. The data are again fitted by a decaying fitting function $P^{th}(t)$
(full line in Fig.~\ref{pTb12}), yielding a mean lifetime $\tau
=\int_{0}^{\infty}t\left( - dP^{th}/dt\right) dt=(64 \pm 5)$ ms.

This significant improvement in the mean trapping time, provided by the LG$%
_{0,12}$ FORT mode with respect to the LG$_{0,1}$ one, is due to the
strongest radial confinement achieved (see Fig.~\ref{compa1-12}). The atom
stays closer to the radial potential minimum where cooling conditions are
optimized and the probability to leave the cavity decreases. This is a
further argument showing that the main advantage of using a doughnut FORT
mode instead of a TEM$_{0,0}$ FORT mode for atom trapping in the case of equal
Stark shifts is just the stronger radial confinement which makes easier to
optimize the cooling conditions. The more the atom is radially confined, the
longer is trapped.

Here we have decreased the width of the radial well by choosing a higher
order doughnut mode and by simultaneously adjusting its waist, so to have an
unchanged radial equilibrium position $\rho _{max}$. This solution may be
practically difficult to implement because one would need different cavity
mirrors for this high-order doughnut FORT mode. Anyway, even though the
implementation of the trapping scheme with a higher-order doughnut mode is
difficult, the latter numerical results clearly show the importance of the
radial confinement, and the direction one has to follow in order to increase
the trapping time of neutral atoms in cavities in the case of equal Stark
shifts induced by the FORT mode.

A possible solution to overcome the practical difficulties of a LG$_{0,12}$ 
experimental set-up can be the following one, provided that one has
enough power and stability for the laser driving the FORT mode.
The same radial potential around $\rho _{\max }$ as that created by
the LG$_{0,12}$ FORT
mode can be achieved using a very intense LG$_{0,1}$ FORT mode.
One can take $S_{0}=4\pi e$ MHz/$\mu$m$^{2}$, leading to $S_{max
}=2\pi \times 400$ MHz, and leave all the other parameters unchanged (see
IV.B for the parameters of the LG$_{0,1}$ FORT mode in case (b)).
In this new configuration, both the radial and the axial wells are 
8 times deeper, and therefore, together with
the desired strong radial confinement 
(substantially the same as that of LG$_{0,12}$ FORT mode), 
we have a strong axial confinement.
Both these facts lead to a considerable increase of trapping time, up to
$2$ seconds (a result of the same order as the one recently
reported in \cite{Buck}),
as demonstrated by the numerical simulation of some trajectories which we
show in the following table: 
\begin{equation}
\begin{tabular}{ccc}
trapping time $\tau $ $(ms)$ & \ \ \ \ \ \ eq. temp. $v_{x}^{rms}$ $%
(cm/s)$ & \ \ \ \ \ \ \ \ type of escape \\ 
157 & 29.4 & \ \ \ \ \ axial \\ 
432 & 20.2 & \ \ \ \ \ axial \\ 
556 & 16.1 & \ \ \ \ \ axial \\ 
684 & 13.7 & \ \ \ \ \ \ radial \\ 
1107 & 10.5 & \ \ \ \ \ \ radial \\ 
1108 & 11.4 & \ \ \ \ \ \ radial \\ 
1177 & 10.7 & \ \ \ \ \ \ radial \\ 
2141 & 10.4 & \ \ \ \ \ \ radial
\end{tabular}
\label{table1}
\end{equation}
According to table (\ref{table1}), the atom is so strongly confined axially
that in some cases (those with the longer trapping times), it leaves
the cavity radially, due to the effect of spontaneous emission.
In these cases, an additional radial cooling
mechanism would be helpful as, for instance, that achieved
via an additional transverse free-space cooling (see \cite{Buck}) or
via an active feedback on radial motion \cite{feedback}.

\section{Conclusions}

We have investigated trapping of single atoms in high finesse optical
cavities using both a quasi-resonant cavity QED field and an intense
classical FORT mode. In particular we have considered the case of doughnut
FORT modes, and we have compared them to the most common case of a fundamental
Gaussian FORT mode (see for example the experiment of Ref.~\cite{Ye-exp}).
Performing full 3-D numerical simulations of the quasi-classical
center-of-mass motion of the atom, we have shown that a doughnut FORT mode
is more suitable than a fundamental 
Gaussian FORT mode to trap the atom, in the case
when the FORT mode shifts both the excited and the ground state down. This
happens when the FORT mode is red-detuned in such a way that the excited
state is relatively closer to resonance with a higher-lying excited state
than with the ground state. This case, even though more difficult to realize
than the standard two-level case where the two Stark shifts are opposite, is
of particular interest for quantum information applications of cavity-QED
systems, where it is important to trap the atom at a given position,
independently of its internal state, so that the quantum manipulation of the
internal state can be easily performed. The advantage of using a doughnut
FORT mode instead of a TEM$_{0,0}$ one is due to the stronger radial
confinement, achieved at a nonzero distance from the cavity axis. In this
way the FORT mode is more suitable to counteract the unavoidable radial
departure caused by the centrifugal potential and by diffusion, by keeping
the atom close to its equilibrium position where the cooling provided by the
quantized cavity mode is optimal.

\section{Acknowledgments}

This work has been partially supported by the European Union through the IHP
program ``QUEST''.


\begin{references}
\bibitem{berm}  {\it Cavity Quantum Electrodynamics}, Advances in Atomic,
Molecular and Optical Physics, Supplement 2, edited by P. Berman (Academic,
New York, 1994)

\bibitem{turch}  Q. A. Turchette, C. J. Hood, W. Lange, H. Mabuchi and H. J.
Kimble, Phys. Rev. Lett. {\bf 75}, 4710 (1995).

\bibitem{varcoe}  B. T. H. Varcoe, S. Brattke, M. Weidinger, H. Walther,
Nature {\bf 403}, 743 (2000).

\bibitem{nog}  G. Nogues, A. Rauschenbeutel, S. Osnaghi, M. Brune, J. M.
Raimond, S. Haroche, Nature {\bf 400}, 239 (1999).

\bibitem{pell}  T. Pellizzari {it et al.}, Phys. Rev. Lett. {\bf 75}, 3788
(1995); J. I. Cirac, P. Zoller, H. J. Kimble, and H. Mabuchi, Phys. Rev.
Lett. {\bf 78}, 3221 (1997); S. J. Van Enk, J. I. Cirac, and P. Zoller ,
Phys. Rev. Lett. {\bf 78}, 4293 (1997); Science {\bf 279}, 205 (1998).

\bibitem{hood}  C. J. Hood, T. W. Lynn, A. C. Doherty, A. S. Parkins, and H.
J. Kimble, Science {\bf 287}, 1447 (2000).

\bibitem{pinkse}  P. V. H. Pinkse, T. Fischer, P. Maunz, and G. Rempe,
Nature {\bf 404}, 365 (2000).

\bibitem{Ye-exp}  J. Ye, D. W. Vernooy, and H. J. Kimble, Phys. Rev. Lett. 
{\bf 83}, 4987 (1999).

\bibitem{ACM}  H. Mabuchi {\it et al.} Opt. Lett. {\bf 21}, 1393 (1996); C.
J. Hood {\it et al.} Phys. Rev. Lett. {\bf 80}, 4157 (1998); P. M\"{u}%
nstermann, T. Fischer, P. Maunz, P. W. Pinkse, and G. Rempe, {\it ibid.} 
{\bf 82}, 3791 (1999).

\bibitem{older}  T. W. Mossberg, M. Lewenstein, and D. J. Gauthier, Phys.
Rev. Lett. {\bf 67}, 1723 {1991}; M. Lewenstein and L. Roso, Phys. Rev. A 
{\bf 47}, 3385 (1993); J. I. Cirac, M. Lewenstein, and P. Zoller, Phys. Rev.
A {\bf 51}, 1650 (1995); A. C. Doherty, A. S. Parkins, S. M. Tan, and D. F.
Walls, {\it ibid.} {\bf 56}, 833 (1997).

\bibitem{ritsch}  P. Horak, G. Hechenblaikner, K. M. Gheri, H. Stecher, and
H. Ritsch, Phys. Rev. Lett. {\bf 79}, 4974 (1997); G. Hechenblaikner, M.
Gangl, P. Horak, and H. Ritsch, Phys. Rev. A {\bf 58}, 3030 (1998); P.
Domokos, P. Horak, and H. Ritsch, J. Phys. B: At. Mol. Opt. Phys. {\bf 34}.
187 (2001).

\bibitem{vule}  V. Vuleti\'{c} and S. Chu, Phys. Rev. Lett. {\bf 84}, 3787
(1999); V. Vuleti\'{c}, H. W. Chan and A. T. Black, Phys. Rev. A {\bf 64},
033405 (2001).

\bibitem{doherty}  A. C. Doherty, T. W. Lynn, C. J. Hood, and H. J. Kimble,
Phys. Rev. A {\bf 63}, 013401 (2001).

\bibitem{feedback}T. Fischer, P. Maunz, P. W. H. Pinkse, T. Puppe, and
G. Rempe, Phys. Rev. Lett. {\bf 88}, 163002 (2002).

\bibitem{vanenk}  S. J. van Enk, J. McKeever, H. J. Kimble, and J. Ye, Phys.
Rev. A {\bf 64}, 013407 (2001).

\bibitem{cohen}  J. Dalibard and C. Cohen-Tannoudji, J. Phys. B: At. Mol.
Opt. Phys. {\bf 18} 1661 (1985).

\bibitem{siegman}  A. Siegman. {\it Lasers}. University Sciences, Mill
Valley (1986)

\bibitem{FORT}  J. D. Miller, R. A. Cline, and D. J. Heinzen, Phys. Rev. A 
{\bf 47}, R4567 (1993).

\bibitem{frequencies}This is possible for Cs atoms in correspondence
with FORT wavelengths from 920 $nm$ upward, as shown in \cite{singapore}
where, in particular, the special wavelength $\lambda =952nm$ causes an
identical shift for the levels. Recently also the 'magic' wavelength $%
\lambda =935nm$ has been used in \cite{Buck}

\bibitem{singapore}  H. J. Kimble, C. J. Hood, T. W. Lynn, H. Mabuchi, D. W.
Vernooy, and J. Ye, {\it Laser Spectroscopy XIV}, Eds. R. Blatt, J. Eschner,
D. Leibfried, and F. Schmidt-Kaler, World Scientific (Singapore, 1999).

\bibitem{Buck}J. McKeever, J. R. Buck, A. D. Boozer, A. Kuzmich, H. C.
Nagerl, D. M. Stamper-Kurn, and H. J. Kimble, quant-ph/0211013 v1 (2002).

\bibitem{Savard}  T. A. Savard , K. M. O'Hara, and J. E. Thomas, Phys. Rev.
A {\bf 56}, R1095 (1997).

\bibitem{gardiner}  C. W. Gardiner, J. Ye, H. C. Nagerl, and H. J. Kimble,
Phys. Rev. A {\bf 61}, 045801 (2000).

\bibitem{Javanainen}  J. Javanainen and S. Stenholm, Appl. Phys. {\bf 21},
35 (1980).

\bibitem{Risken}  H. Risken, {\it The Fokker-Planck Equation}, 2$^{nd}$
Edition (Springer, 1989).

\bibitem{gardiner2}  C. W. Gardiner, {\it Handbook of Stochastic Methods for
Physics, Chemistry and the Natural Sciences}, 2$^{nd}$ Edition
(Springer-Verlag, Berlin, 1985).

\bibitem{Tan}  S. M. Tan, J. Opt. B {\bf 1}, 424 (1999). S. M. Tan's Quantum
Optics Toolbox at www.phy.auckland.ac.nz/Staff/smt/qotoolbox/download.html.

\bibitem{Vernooy}  D. W. Vernooy and H. J. Kimble, Phys. Rev. A {\bf 56},
4287 (1997).

\bibitem{hotatoms}L.-M. Duan, A. Kuzmich, and H. J. Kimble,
quant-ph/0208051 (2002).

\bibitem{kaleidoscopio}P. Horak, H. Ritsch, T. Fischer, P. Maunz, T.
Puppe, P. W. H. Pinkse, and G. Rempe, Phys. Rev. Lett. {\bf 88}, 043601
(2002).

\bibitem{Scaling}P. Horak and H. Ritsch, Phys. Rev. A {\bf 64}, 033422 
(2001).

\end{references}
\end{document}